\documentclass[aps,prd,floats,nofootinbib,preprintnumbers,superscriptaddress]{revtex4}
\usepackage{epsfig}
\usepackage{amsmath}
\usepackage{amssymb}
\usepackage{amsthm}

\begin{document}

\preprint{NUHEP-TH/06-03}

\title{What can we learn from neutrino electron scattering?}

\author{Andr\'e de Gouv\^ea}
\affiliation{Northwestern University, Department of Physics \& Astronomy, 2145 Sheridan Road, Evanston, IL~60208, USA}

\author{James Jenkins}
\affiliation{Northwestern University, Department of Physics \& Astronomy, 2145 Sheridan Road, Evanston, IL~60208, USA}

\begin{abstract}
Precision tests of the standard model are essential for constraining
models of new physics.  Neutrino--electron elastic scattering offers
a clean probe into many electroweak effects that are complimentary
to the more canonical measurements done at collider facilities. Such
reactions are rare, even as compared with the already tiny
cross-sections for neutrino--nucleon scattering, and competitive
precision measurements have historically been challenging to obtain.
Due to new existing and proposed high-flux neutrino sources, this is
about to change.  We present a topical survey of precision
measurements that can be done with neutrino--electron scattering in
light of these new developments.  Specifically, we consider four
distinct neutrino sources: nuclear reactors, neutrino factories,
beta-beams, and conventional beams.  For each source we estimate the
expected future precision of several representative observables,
including the weak mixing angle, neutrino magnetic moments, and
potential leptonic $Z^{\prime}$ couplings.  We find that future
neutrino--electron scattering experiments should add non-trivially
to our understanding of fundamental physics.

\end{abstract}

\maketitle

\setcounter{equation}{0} \setcounter{footnote}{0}
\section{Introduction}
\label{sec:intro}

Neutrino--electron scattering offers a clean probe into the standard
model of particle physics as well as many of its extensions.
`Clean' refers to the fact that this process is very well
understood theoretically.  There are no hadronic complications, so that the
underlying electroweak physics (including potential deviations from standard model
expectations) is directly accessible.  It is therefore ``easy''
to use such reactions to test the consistency of the standard model (SM),
determine precision electroweak parameters, and look for signatures
of new physics.  When this is considered along with the large
variety of natural and artificial sources that yield high-flux
neutrino samples of multiple flavors over a vast energy range, it
seems that neutrino--electron scattering provides an ideal laboratory
for electroweak studies.

This line of reasoning appears misleading when realistic considerations are made.
While theoretically ideal, the study of neutrino--electron scattering is
experimentally challenging due to its tiny cross-section, which forces
one to pursue very intense sources and large targets.
More serious is the fact that  the neutrino--nucleon scattering cross-section
is generally three orders of magnitude larger
and serves as a large potential source of background.  Naively,
one may consider simply subtracting off such background statistically.
 In this case, the uncertainty induced
by the subtraction is approximately $\sqrt{10^3} \approx 30 $ times
larger than the intrinsic statistical uncertainty of the signal;
clearly unacceptable for precision measurements.  Of course,
this is a worst case scenario: experimental set-ups usually allow one to
isolate the signal events by performing various ``cuts'' on the data.
Exactly how this is done depends on the energy and flavor of the
incident neutrino beam, as well as on the details of the detector.

The signal  we concentrate on is a single forward electron with no other detector
activity, and all detectors considered can distinguish, with varying degrees of
success, electrons from various potential background sources.  Cuts
on event timing, rate of energy loss in the detector,
and threshold energy, to name a few, aid in this endeavor (details are discussed in
Sec.~\ref{sec:fluxes}).  Even with optimistic particle identification
abilities, experiments must still account for irreducible backgrounds.
This is particularly relevant in $\nu_e$ and $\bar{\nu}_e$ sources,
where the charged current
neutrino--nucleon reaction $\nu_e N \rightarrow e X$ can yield a final state that is
often consistent with a single recoil electron. Such backgrounds, however,
can still be reduced by exploiting additional constraints.  For example,
electrons produced by neutrino--electron scattering
are constrained by kinematics to have small transverse
momenta $p_t \propto \sqrt{m_e}$, whereas the electron $p_t$ distribution in most
background events is much broader, $p_t\propto\sqrt{m_{\rm nucleon}}$.
Therefore, an experiment with good $p_t$ resolution can significantly constrain this class of
background events.
 After all available data analyzing resources are spent, one is
 (hopefully) left with a signal-dominated event sample.  This being the case,
 the remaining beam-related background can be modeled and
subtracted, inducing a (much smaller) statistical uncertainty on
the final data sample.  Backgrounds related to other neutrino/radiation sources originating
from extra-terrestrial, terrestrial, and artificial origins are
typically controlled by introducing shielding and imposing clever
$p_t$, timing, and energy cuts.  The resulting backgrounds and
uncertainties will typically be small, and are therefore not
considered here. Finally, one must also account for other
experimental systematic uncertainties in the final analysis, which may or may not dominate
the sensitivity budget.

These considerations imply that measurements of neutrino--electron
elastic scattering can be limited mainly by statistics and
uncertainties related to the neutrino source.
In order to accumulate enough statistics, one is required
to commit to long running
experiments with large detectors close to the neutrino source; a
practice that has proven fruitful in the recent past, but not sufficient to
yield results competitive with other tests of electroweak physics.
Significant progress is expected to be made with the advent of improvements to the neutrino sources
themselves, both in statistics and flux-normalization.

Currently existing sources that could be used for neutrino--elastic scattering purposes can be
broadly classified as either reactors or conventional neutrino beams, and are both
discussed in Sec.~\ref{sec:fluxes}. These are still subject to some of the
limitations described above, which, in some cases, prevent
competitive electroweak ``precision'' measurements. Recent progress in the
development of two new classes of neutrino beams inspires the
possibility of sidestepping these limitations and thereby testing
the SM to unparalleled accuracy. Neutrino factories and
$\beta$-beams offer high luminosity neutrino beams with well-known energy
spectra  \cite{bbeam_nufact_RD}. Here, we explore the potential of
neutrino--electron scattering experiments in light of our enhanced knowledge of
these sources.

The paper is organized in the following manner.  Section
\ref{sec:formalism} reviews the relevant tree-level SM
cross-sections, making brief mention of first order electroweak
corrections.  We then describe the various neutrino sources --
reactors, conventional beams, $\beta$-beams and $\nu$-factories -- used in
our analysis. In each case we describe the energy spectrum, as well as
uncertainties and backgrounds relevant to their associated neutrino--electron
scattering experiments.  Section \ref{sec:Results} begins
with a short description of our analysis, after which we review and
motivate various observables and present our results on projected
sensitivities to each within the context of future scattering
experiments. Specifically, we discuss measurements of the weak
mixing angle $\theta_W$, neutrino electromagnetic moments $\mu_\nu$,
neutrino neutral current left-handed couplings $\rho$, and potential
leptonic $Z^{\prime}$ couplings. We conclude in Section~\ref{sec:end} with a
summary of our results and an outlook for the future.

\setcounter{footnote}{0}
\setcounter{equation}{0}
\section{Formalism}
\label{sec:formalism}

We are interested in ``elastic'' neutrino--electron and antineutrino--electron scattering,
characterized by $\nu_{\ell}e^-\to\nu_{\ell'}e^{-}$, where $\ell,\ell'=e,\mu,\tau$ and
$\nu$ stands for either a neutrino or an antineutrino state. Note that, given our
inability to identify the neutrino flavor after it has scattered off the target electron, there is no
way of recognizing whether the scattered neutrino has the same lepton-flavor number
or lepton number as the incoming one.

The basis of this analysis is the differential event spectrum
${\rm d}N(T)/{\rm d}T$.  This is the number of neutrino--electron elastic
scattering events within the interval $T$ to $T+{\rm d}T$ of electron recoil
kinetic energy.  It involves the convolution of the differential
cross-section ${\rm d}\sigma(T,E_\nu)/{\rm d}T$ and the incoming neutrino energy spectrum,
${\rm d}\Phi(E_\nu)/{\rm d}E_\nu$.  More than one neutrino
flavor/helicity may be produced at each of the sources listed in
Sec.~\ref{sec:fluxes} and, since the final state electrons scattered
from the various neutrino types are experimentally
indistinguishable, their contributions must be incoherently added, leading to:

\begin{equation}
\frac{{\rm d}N(T)}{dT} = {\rm (time)} \times {\rm (\# targets)} \times
\sum_i^{{\rm flavors}} \int {\rm d}E_\nu
\frac{{\rm d}\Phi_i(E_\nu)}{{\rm d}E_\nu} \frac{{\rm d}\sigma_i(T,E_\nu)}{{\rm d}T},
 \label{eqn:rate_spectrum}
\end{equation}
where (\# targets) is the
total number of target-electrons in the detector and (time) is the time duration
of the experiment.  In order to use Eq.~(\ref{eqn:rate_spectrum}), we
must know the flux and the cross-sections, along with their associated
uncertainties.  These are reviewed in the following subsections.  For the
remainder of this work, unless stated otherwise, we assume a $100$~ton detector of similar capabilities as
Miner$\nu$a  \cite{minerva}, located $100$~m from the neutrino
source, running for one year.\footnote{One year is defined to be $3.16\times 10^7\,{\rm s}$.}

\subsection{Cross sections}
\label{subsec:cross}

 In the standard model (SM), all tree-level differential
cross-sections for neutrino--electron scattering can be expressed as
\begin{equation}
\frac{d\sigma}{dT}(\nu_{\ell} e\rightarrow \nu_{\ell} e) = \frac{2G_\mu^2m_e}{\pi
E_\nu^2}\left[a^2E_\nu^2 + b^2(E_\nu - T)^2 - abm_eT\right],
 \label{eqn:dsig_dT_SM}
\end{equation}
where $G_\mu$ is the Fermi constant, $E_\nu$ is the energy of the
incident neutrino and $T$ is the kinetic energy of the recoil
electron.  $a$ and $b$ are process-dependent constants that, within
the SM, depend on the  weak mixing angle
$\theta_W$, as tabulated in Table~\ref{table:couplings}.  The cross term proportional to $m_e$, the mass of the
electron, is relevant  for low energy
applications, but is  negligible in processes where
$E_\nu \gg m_e$.

\begin{table}[ht]
\centering \caption{Standard model $a$ and $b$ parameter values for
the differential cross-section, given by Eq.~(\ref{eqn:dsig_dT_SM}).
Here $s^2 \equiv \sin^2\theta_W \approx 0.23149 \pm 0.00015$
\cite{pdg} where $\theta_W$ is the electroweak mixing angle, and
$\ell=\mu,\tau$.\label{table:couplings} }
\begin{tabular}{|c||c|c|c|c|} \hline
   & $\nu_e e\rightarrow \nu_e e $& $\bar{\nu}_e e\rightarrow \bar{\nu}_e e$ & $\nu_\ell e \rightarrow \nu_\ell e$ & $\bar{\nu}_\ell e\rightarrow \bar{\nu}_\ell e$\\
   \hline
  $a$~ &$-\frac{1}{2} - s^2$  & $-s^2$ & $\frac{1}{2} - s^2$ & $-s^2$ \\ \hline
  $b$~ & $-s^2$ & $-\frac{1}{2} - s^2$& $-s^2$ & $\frac{1}{2} - s^2$ \\\hline
\end{tabular}
\end{table}

Since the incident
neutrinos are all produced by some charged current process, we assume that
all incoming neutrinos  (antineutrinos) are
strictly left-handed (right-handed). Given our understanding of the charged current
interactions, this is an excellent
approximation even if one considers the existence of exotic helicity-flipping processes
whose amplitude are necessarily
proportional to the neutrino mass, and therefore negligible for all practical purposes.
We further assume that all electron targets are unpolarized.

For $\ell =
\mu,\tau$, the scattering process proceeds via t-channel $Z$-boson exchange.
%
The $\nu_e e \rightarrow \nu_e e$ and
$\bar{\nu}_e e \rightarrow \bar{\nu}_e e$ reactions proceed via a
combination of t-channel $Z$-boson, and t/s-channel $W$-boson exchange,
respectively, and are related by $a\leftrightarrow b$
exchange.
In Sec.~\ref{sec:Results}D, we will discuss the sensitivity of neutrino--electron
scattering to the left-handed neutrino coupling to the $Z$-boson, referred to as $\rho$. In the SM,
$\rho=1$ at tree-level. For arbitrary values of $\rho$, the cross section for $\nu_{\ell}e^-$ scattering simply scales
with $\rho^2$ in the $\ell=\mu,\tau$ case, while the dependency is more involved in the $\ell=e$ case.
We return to this issue in Sec.~\ref{sec:Results}D.

Current experimental precision allows the extraction of many
electroweak observables to better than $1\%$, introducing the need
to go beyond the simple tree-level cross-sections outlined above.
Indeed, neutrino electron scattering experiments must include full
first (and perhaps second) order corrections into their analysis to
maintain consistency  \cite{Bahcall_radCorr}.  These
corrections are theoretically well-known, and can be easily applied
to data analysis  \cite{Marciano_radCorr,Sarantakos_radCorr}. See
 \cite{practical_radCorr} for a pedagogical review of calculations
involving electroweak radiative corrections.  Here we briefly
summarize the results of such first order effects, utilizing the
minimal subtraction ($\overline{MS}$) renormalization scheme. See
 \cite{nu_eReview,Bahcall_radCorr} for the full expressions.

The full $\mathcal{O}(\alpha)$ corrections to the $\nu e$
cross-sections given by Eq.~(\ref{eqn:dsig_dT_SM}) involve one loop
effects as well as photon bremsstrahlung.  QED effects (in the
relevant high energy limit) are well described by a $T$ dependent,
$\mathcal{O}(\alpha)$ modification of the $a,b$ parameters:
\begin{equation}
a^2 ({\rm or}~b^2) \rightarrow a^2({\rm or}~b^2) \left[1+ \alpha F_{a(b)}(T)\right],
\end{equation}
where $F_a$ and $F_b$ are dimensionless functions of $T$.
The remaining corrections are generally $q^2$ dependent and
parameterized by the running of the weak mixing angle
$\sin^2\theta_W$, and the deviation of $\rho$ from its tree-level
value of $1$.  These relations depend weakly on the Higgs mass,
which we take to be 150~GeV.  The
net result is an $\mathcal{O}(5\%)$ shift in the differential
cross-section, integrating to an approximately $1\%$ effect on the
total cross-section.

Given the precision with which next-generation experiments can probe
the physics of neutrino--electron scattering, higher order corrections,
while small, are by no means irrelevant, and need to be taken into account.
We would like to emphasize, however, that the effect of higher
order SM corrections is negligible \emph{for the
purposes of this paper}. This is a consequence of the fact that we are
interested in gauging the \emph{precision} ($\delta o$, $o$ for observable)
of various measurements,
not in computing what their central values ($\bar{o}$ for the extracted value) are.
Generally, $\frac{\delta
o}{\bar{o}} \approx~\rm few\%$ for the observables of interest, thus the
dependence on electroweak corrections $\delta_{EW}\left(\frac{\delta
o}{\bar{o}}\right) \approx (1\%)^2 = 0.01\%$ can be ignored.  That
being said, our analyses do in fact incorporate
first order SM effects. However, in the spirit of simplicity, we
shall refrain from mentioning them and always refer to the tree-level
cross-sections of Eq.~(\ref{eqn:dsig_dT_SM}) when necessary.

\subsection{Fluxes}
\label{sec:fluxes}

\begin{figure}
\begin{center}
\includegraphics[scale=0.9]{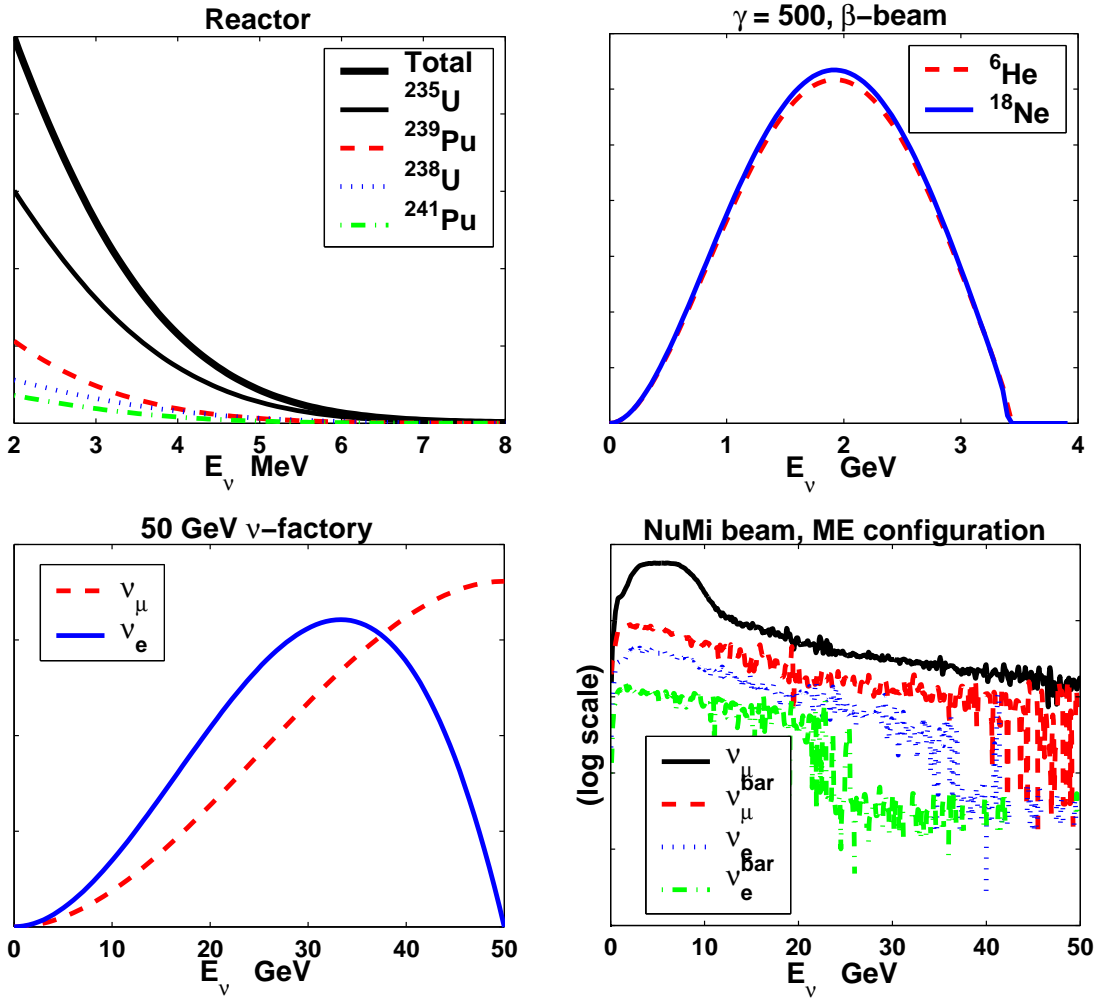}
\caption{Neutrino energy spectra for different neutrino sources.
Flux normalizations are arbitrary. }\label{figure:fluxes}
\end{center}
\end{figure}

In our analysis, the specific details of the incoming neutrino
energy spectrum matter little for determining an experiment's
sensitivity to an observable.  How well we know that spectrum is,
however, of the utmost importance, as is the overall luminosity, mean
beam energy, and neutrino flavor composition.  The sample
sources used in our analyses were chosen to span a large range of
possible configurations.  Specifically,
we consider four types of neutrino sources yielding distinct flavor
content and energy spectra over a broad
energy range. These are nuclear reactors, neutrino factories, $\beta$-beams
and conventional beams.
Their respective energy spectra are depicted in Fig.~\ref{figure:fluxes}. Note that there
are other potential experimental setups capable of precision neutrino--electron scattering
\cite{meas_nue_scat}. These will not be considered in this study.

\subsubsection{Reactor}

Nuclear reactors are intense sources of low energy ($< 10{\rm MeV}$)
$\bar{\nu}_e$'s, and continue to play a central role in neutrino
physics  \cite{reactor_review}.  The nuclear fuel in a modern light
water reactor is typically enriched with $(2-5)\%$ $^{235}{\rm U}$
and produces 3~GW of thermal power. On average, 200~MeV
and $6$ neutrinos are released with each fission, thus yielding around
$5\times10^{20} \bar{\nu}_e/{\rm s}$  \cite{reactor_review}.
The main limitation of a reactor as a source of anti-neutrinos is
that the fissions occur at rest, implying isotropic emission: you can't ``focus''
the reactor beam.

While it is relatively simple to predict the absolute magnitude of emitted
neutrinos, predicting their energy spectra requires detailed
modeling. In an actual neutrino experiment, the thermal power of the
reactor\footnote{Other reactor quantities such as water
pressure/temperature must be measured and used in the modeling of
the neutrino spectrum.  These affect the spectrum weakly.
See  \cite{reactor_review} for more details.}
is measured as a function of time.  With this and the initial fuel
composition, the fission rate can be modeled for the dominant
isotopes.  These are $^{235}{\rm U}$, $^{239}{\rm Pu}$, $^{238}{\rm
U}$, and $^{241}{\rm Pu}$, and they typically fission, in an average
fuel cycle, by the proportions $25:10:4:3$.  Other isotopes such as
$^{240}{\rm Pu}$ and $^{242}{\rm Pu}$ contribute to the flux at the
$\mathcal{O}(0.1\%)$ level and can be safely neglected. Combined with
knowledge of the induced decay of these dominant isotopes by thermal
neutrons, this procedure predicts the reactor $\bar{\nu}_e$ energy
spectrum to better than $10\%$ and the total flux to nearly
$1\%$  \cite{low_Ereactor}. Here, we will employ
an empirical relation to describe the neutrino
flux  \cite{reactor_param1,reactor_param2}.\footnote{The chemical composition of the fuel in the reactor
also varies as a function of time, and must be considered in order to translate reactor flux measurements into physics observables.
The impact of this time-evolution is, however, negligible when it comes to estimating the precision with which physics observables can be measured.
Therefore, for the purposes of the upcoming analyses, it suffices to deal with the average fuel compositions.}
\begin{equation}
\frac{d\Phi_{\bar{\nu}_e}(E_{\nu})}{dE_{\nu}} = N\sum_{i}n_i e^{a_{0i} + a_{1i}E + a_{2i}E^2},
\label{eqn:reactor_flux}
\end{equation}
where $i$ runs over the dominant parent fission isotopes $^{235}{\rm
U}$, $^{239}{\rm Pu}$, $^{238}{\rm U}$, and $^{241}{\rm Pu}$ and the
$n_i$'s are related by the proportions mentioned above.  Table
\ref{table:reactor_flux} lists the coefficients of the exponentiated
polynomial in Eq.~(\ref{eqn:reactor_flux}) for each isotope \cite{prec_reactor_nueb}, as well
as the relative $n$ values \cite{reactor_review}.  The top-left
panel of Fig.~\ref{figure:fluxes} depicts them individually along
with the total $\bar{\nu}_e$ flux.  The absolute normalization $N$ is
found from the requirement that the total neutrino rate be
$2\times10^{20} ~( \rm Power/GW)$~s$^{-1}$.

\begin{table}[t]
  \centering
   \caption{Coefficients of the exponentiated second order polynomial in the reactor anti-neutrino flux Eq.~(\ref{eqn:reactor_flux}), adapted from  \cite{prec_reactor_nueb}. Average $n_i$ values were extracted from \cite{reactor_review}.}
  \label{table:reactor_flux}
  \begin{tabular}{|c|cccc|}
  \hline
  $i$ & $a_{0i}$ & $a_{1i} MeV^{-1}$ & $a_{2i} MeV^{-2}$ & $n_i$ \\
  \hline
  $^{235}{\rm U}$  & $0.904$ & $-0.184$ & $-0.0878$ & $25$  \\
  $^{239}{\rm Pu}$ & $1.162$ & $-0.392$ & $-0.0790$ & $10$ \\
  $^{238}{\rm U}$ & $0.976$ & $-0.162$ & $-0.0790$  & $4$ \\
  $^{241}{\rm Pu}$ & $0.852$ & $-0.126$ & $-0.1037$ & $3$ \\
  \hline
  \end{tabular}
 \end{table}

At these energies, the inverse $\beta$-decay process
\begin{equation}
\bar{\nu}_e p \rightarrow e^+ n
\label{eqn:inverse_beta_decay}
\end{equation}
is the dominant source of background by a factor greater than $100$.
Due to its threshold of $1.8{\rm MeV}$, only about $25\%$ of the
released neutrinos will trigger such a reaction.  This begs the
question of why neutrino scattering experiments are done with the
high energy $>2$~MeV tail of the spectrum, in the presence of
lower statistics and a significant source of background.  Some
experiments do, in fact, use low energy reactor neutrinos; they are
primarily designed to search for neutrino magnetic moments or study
neutrino-nucleus coherent scattering \cite{coherent}.  The difficulty in working
within this energy range is the uncertainty in the neutrino flux,
which can be as large as $30\%$  \cite{low_Ereactor}.  Flux
measurements have not been made below $\approx 2{\rm MeV}$ and
theoretical calculations are not reliable due to an increase in the
number of $\beta$-decay chains with low $Q^2$ values, many of which are
not completely understood.  To complicate matters further,
long-lived isotopes, residing in spent fuel stored on-site, radiate
in this range and must be tracked and accounted for in a reliable
analysis.

In the $(2-8)$~MeV region, a detector capable of distinguishing the
inverse $\beta$-decay reaction from the signal with high efficiency
is needed. A study of the uncertainties associated with reactor
neutrinos as relevant to neutrino--electron scattering at these
energies was presented in  \cite{Conrad_reactor}.  There, the
authors assume a 26.5~ton CHOOZ-like detector  \cite{CHOOZ} composed
of oil scintillator located $\sim 225$~m from two $3.6$~GW nuclear
reactors. Such an experiment is optimized for $\bar{\nu}_e p
\rightarrow e^+ n$ detection, but can also be used to identify and
reconstruct the energy of the final state electron in $\bar{\nu}_e e
\rightarrow \bar{\nu}_e e$.  In general, these two processes are
distinguished by the detection of the final state neutron in
Eq.~(\ref{eqn:inverse_beta_decay}) via its capture on ${\rm H}$ and
${\rm Gd}$ nuclei after a characteristic time in which the neutron
thermalizes.  This procedure induces some systematic uncertainty on
background subtraction from the failure to identify some neutrons
within the characteristic time window, resulting in the
misidentification of some background as signal. Additionally, other
sources of uncertainty arising from experimental factors such as
energy calibration and efficiencies must also be included for a
realistic analysis.  Throughout this work we refer to these simply
as systematic effects and treat them as we do the background
subtraction uncertainty.  Backgrounds unrelated to the source are
controlled by shielding and comparing the on/off reactor states
needed for refueling. Following
 \cite{Conrad_reactor}, the total uncertainty from background can be
minimized to the $1\%$ level by utilizing various experimental cuts.
Of particular relevance to our analysis, a cut on the electron
visible energy $3{\rm MeV} < T < 5{\rm MeV}$ must be applied to
achieve such precision. Additionally, by normalizing to the inverse
$\beta$-decay sample, an uncertainty of order $0.1\%$ can be achieved in
the overall neutrino flux normalization \cite{Conrad_reactor}, an order of magnitude
better than that achieved from reactor modeling alone.

Future reactor experiments designed to operate in our selected energy window are optimized to search for the elusive neutrino mixing angle $\theta_{13}$ \cite{reactor_whitepaper} (see table 1 of \cite{reactor_future} for a concise list of future experiments along with many of their projected specifications such as reactor power, detector baselines, and fiducial mass).   All of the proposed facilities include at least one near detector at around 100~m from the source to help control the various source-related systematic uncertainties.  Such sites can serve as ideal next-generation laboratories for the study of low energy neutrino--electron scattering.  Of these, the Double CHOOZ experiment \cite{DCHOOZ} should be the first to begin taking data, and could significantly help explore many of the topics surveyed here.

For our analyses we assume a single 3~GW reactor, with a flux
given by  Eq.~(\ref{eqn:reactor_flux}) known to $0.1\%$.
Furthermore, we apply the visible energy cut $3{\rm MeV} < T < 5{\rm
MeV}$ described above, along with an induced $1\%$ systematic
uncertainty arising from background subtraction.  With this in
place, an experiment running for one year should record
approximately $10^4$ signal events, which implies a $1\%$
statistical uncertainty.

\subsubsection{Neutrino Factory}

The concept of a neutrino factory has received much attention in
recent years and is now entering a serious development
stage  \cite{bbeam_nufact_RD}.  The concept is simple:  produce and
isolate a copious amount of muons from an intense ($>1{\rm MW}$)
proton beam incident on a fixed target.  Boost them to the desired
energy and then inject them into a storage ring with a long straight
section. The boosted muons in the straight section will decay in
flight into two neutrinos nearly $100\%$ of the time  \cite{pdg}:
\begin{eqnarray}
\label{eqn:nufact_mreaction}
\mu^- &\rightarrow& e^- \nu_\mu \bar{\nu}_e, \\
\label{eqn:nufact_preaction}
\mu^+ &\rightarrow&
e^+\bar{\nu}_\mu\nu_e.
\end{eqnarray}
Neutrinos that result from muon decay in straight
sections are beamed in the forward direction by an amount dependent
on the boost factor $\gamma$ of the parent muon.  The specific flavor
composition of the resulting collimated neutrino beam is a simple
consequence of the sign of the selected parent muon.  Clearly, the beam will consist of
equal proportions of muon-type and electron-type neutrinos, where
one is a particle and the other an antiparticle.  The geometry of
the storage ring can be optimized so that a maximum percentage of
muons decay in the straight section (approximately 35\%).

In the muon's rest frame, the decays Eq.~(\ref{eqn:nufact_mreaction}) and Eq.~(\ref{eqn:nufact_preaction})
are described by the following well-known
expressions  \cite{bbeam_nufact_RD}:
\begin{eqnarray}
\label{eqn:nufact_mufluxcm} \frac{d^2\Phi_{\nu_\mu}}{dE_\nu
d\Omega_{cm}} &\propto&
\frac{4E^2_\nu}{\pi m^4_\mu}\left(3m_\mu - 4E_\nu\right),\\
\label{eqn:nufact_efluxcm} \frac{d^2\Phi_{\nu_e}}{dE_\nu
d\Omega_{cm}} &\propto& \frac{24E^2_\nu}{\pi m^4_\mu}\left(m_\mu -
2E_\nu\right),
\end{eqnarray}
where $E_\nu$ refers to the energy of the emitted neutrino in the
muon's rest frame.
Eq.~(\ref{eqn:nufact_mufluxcm}) and Eq.~(\ref{eqn:nufact_efluxcm}) are
valid for both $\mu^+$ and $\mu^-$ decays, provided that the
muon beam has zero net polarization.  For a polarized beam, an
additional term is generated that changes sign under $\mu^+
\leftrightarrow \mu^-$. This possibility is not considered here,
but can easily be introduced and is not expected to affect any of our results
(we refer readers to  \cite{bbeam_nufact_RD} and references
therein for the status and implications of polarized muon beams as
applied to a neutrino factory).

Given
Eq.~(\ref{eqn:nufact_mufluxcm}) and Eq.~(\ref{eqn:nufact_efluxcm}),
it is straightforward to obtain the electron-type and muon-type neutrino fluxes at
any boosted reference frame.
The bottom-left panel of Fig.~\ref{figure:fluxes} shows the on-axis energy
spectra of a 50~GeV neutrino factory beam. Both
the $\nu_\mu$ and $\nu_e$ beam components are shown without
distinguishing between neutrino and anti-neutrino because they yield
the same spectrum. The absolute normalization is obtained by
requiring that the integrated flux over solid angle and energy
yields $10^{20}$~decays/year  \cite{nufact_norm}.
Neutrino factory designs aim at  reaching a $0.1\%$
uncertainty on the flux normalization  \cite{front_end_nufact}.  This
can be further reduced by half from normalizing to the muon
regeneration process $\nu_\mu e \rightarrow \mu^- \nu_e$ and
$\bar{\nu}_e e \rightarrow \mu^- \bar{\nu}_\mu$.  A neutrino factory
running in the $\mu^-$ mode of Eq.~(\ref{eqn:nufact_preaction}) is
subject to a very real source of background via the inverse beta-decay
reaction, which must be reduced by applying $p_t$ cuts.  A
beam running in the $\mu^+$ mode is not affected by this, but still
may be subject to other background events that contain a single
final state electromagnetic shower that mimics the electron signal,
such as coherent and diffractive $\pi^0$ production. These must be dealt with during data analysis.
After all expected analysis cuts, the resulting
signal-to-background ratio is projected to be better than
five  \cite{front_end_nufact}. Under such high statistics
conditions, backgrounds are negligible when compared to the overall flux normalization
uncertainty and additional systematic effects.

For our analyses, we assume $10^{20}$ muon decays per year in the
straight section of the storage ring producing a neutrino flux given
by the boosted version of Eq.~(\ref{eqn:nufact_mufluxcm}) and Eq.
(\ref{eqn:nufact_efluxcm}).  The uncertainty on the absolute
normalization is taken to be $0.1\%$ and $0.05\%$ for $\mu^+$ and
$\mu^-$ beams respectively, with negligible induced statistical
uncertainty arising from background subtraction. Additional
systematic uncertainties in such neutrino factory experiments are
difficult to estimate and may be large as compared to the values
listed here.  We therefore perform our analysis assuming a
systematic uncertainty ranging from $(0 - 5)\%$.  With this, an
experiment running for one year with a 50~GeV beam should record
approximately $10^9$ elastic scattering events, inducing a
statistical uncertainty of only $0.003\%$.

\subsubsection{$\beta$-beam}

$\beta$-beams are newly envisioned facilities that will produce an
intense beam of electron type neutrinos ($\nu_e$ or $\bar{\nu}_e$)
with well-known energy spectra.  These beams are
virtually free of contamination by other neutrino flavors.  The idea
behind a $\beta$-beam is very similar to that of a neutrino
factory  \cite{bbeam_nufact_RD}.  $\beta$-decaying isotopes are produced
by an optimized fixed target collision. They are then
accelerated and placed in a storage ring where they undergo $\beta$-decay,
producing a collimated neutrino beam. Current isotopes of interest
are $^{18}{\rm Ne}$ for $\nu_e$ production and $^6{\rm He}$ for
$\bar{\nu}_e$ production. Approximately $10^{18}$ decays per year
are expected at a $\beta$-beam facility for either nuclei, assuming the
existing design \cite{high_gamma_bbeams}, $35\%$ of which should
occur in the straight section of the storage ring and therefore
constitute the beam.

$\beta$-decay kinematics are well-known and lead to the following
approximate form for the neutrino energy spectrum in the ion's rest frame:
\begin{equation}
\frac{d^2\Phi}{dE_\nu d\Omega_{cm}} \propto E_\nu^2(E_0 -
E_\nu)\sqrt{(E_0 - E_\nu)^2 - m_e^2},
\end{equation}
where $E_0$ is the electron end point energy; 3.5~MeV for
$^6{\rm He}$ and 3.4~MeV for $^{18}{\rm Ne}$.
Hence,
\begin{equation}
\left.\frac{d^2\Phi}{dE_\nu d\Omega_{lab}}\right|_{\theta_{lab}
\approx 0} \propto \gamma^2E_\nu^2(E_m - E_\nu)\sqrt{(E_m - E_\nu)^2
- (2\gamma m_e)^2}, \label{eqn:bbeam_flux}
\end{equation}
where $E_\nu = 2\gamma E_\nu^{cm}$ is now the transformed energy in
the boosted frame and $E_m = 2\gamma E_0$ is the maximum neutrino
energy.  The top-right panel of Fig.~\ref{figure:fluxes} shows the
shape of the $\beta$-beam flux given by Eq.~(\ref{eqn:bbeam_flux}) for
both isotopes mentioned above and a boost factor $\gamma = 500$, assuming the same
overall normalization for each. The small difference between the
curves is due to the differences in the $\beta$-decay end-point energy.
Once again the normalization is found by conditions on the
integrated flux and is assumed, as in the neutrino factory case, to
be known to approximately $0.1\%$.  The proposed boost factors range
from $\gamma = 60$ with a mean energy of 0.2~GeV to $\gamma =
2500$ with a mean energy of 7~GeV.  At these energies, the
primary source of background is quasi-elastic scattering (deep
inelastic processes become dominant at the high energy facilities).
The number of such events can be reduced  by imposing kinematical cuts on
$p_t$.

For our analyses, we assume a $\gamma = 500$ $\beta$-beam source
consisting of $1.1\times 10^{18}$ and $2.9\times 10^{18}$ decays per
year at facilities running in the $\nu_e$ ($^{18}{\rm Ne}$) and
$\bar{\nu}_e$ ($^6{\rm He}$) modes
respectively \cite{high_gamma_bbeams}, with an energy spectrum given by
Eq.~(\ref{eqn:bbeam_flux}) carrying an overall $0.1\%$ uncertainty.
Although the signal-to-background ratio for such $\beta$-beam
experiments should be large enough to neglect the statistical
uncertainty induced from background subtraction, additional
systematic uncertainties may still be large as compared with the
other characteristic uncertainties of the system.  We therefore
perform our analysis assuming systematic uncertainties ranging from
$(0-5)\%$.  Under these conditions, an experiment running for one
year should record $6\times 10^5$ elastic scattering events inducing
a statistical uncertainty of only $0.13\%$.

\subsubsection{Conventional Beams}
\label{sec:conventional_beams}

We define a \emph{conventional beam} broadly as any neutrino source
arising primarily from the decay of accelerator-produced pions or
kaons.
There are currently several conventional neutrino beams  in
operation or in the development stage.  Many were constructed for the primary
purpose of studying long/medium baseline neutrino oscillations, but can also
be used to study neutrino--electron scattering.  A detector, as described
here, placed close to the neutrino source would yield a high neutrino--electron
statistics sample.  This would lead
not only to an enhancement of our knowledge of fundamental particle properties,
but would serve as a source of normalization for the oscillation
experiments as well as reduce uncertainties on cross-sections needed
to extract oscillation parameters.  One example is the
K2K  beamline, originating from the KEK accelerator facility
 in Japan which yields a high
luminosity broadband $\nu_\mu$ beam peaking in the sub-GeV
energy range to the Super-Kamiokande detector 250~km away \cite{k2k,t2k_k2k}.
A more powerful beam at the currently-under-construction J-PARC facility is being planned \cite{t2k_k2k}.
In the USA, Fermilab is currently home to two important
conventional neutrino beams. The booster neutrino beamline provides a low
energy $(0.5-1.5)$~GeV, $\nu_\mu$ beam to the MiniBooNE experiment and
may, in the future, also serve the proposed FINeSSE (Fermilab
Intense neutrino Scattering Scintillator Experiment) experiment with
an overall flux uncertainty of approximately $5\%$ \cite{Finesse}. At
much higher energies, the NuMI (Neutrinos at the Main Injector) beam
is planned to power the Miner$\nu$a \cite{minerva} detector, which is to be located
behind the  MINOS near detector.  The NuMI beam can operate in
different configurations ranging in peak energy from
$(3-15)$~GeV.  Additionally, it has the option of running in
the ``negative" mode, dominated by $\nu_\mu$, or the reverse
``positive" mode, dominated by $\bar{\nu}_\mu$. Planned upgrades to
the Fermilab proton accelerator would significantly enhance the
performance of the NuMI beam as it applies to both neutrino
oscillation and scattering experiments \cite{proton_driver}.

Generally, conventional beams consist of $\nu_\mu$, $\bar{\nu}_\mu$,
$\nu_e$ and $\bar{\nu}_e$, at least to some degree. Muon-type
neutrinos are always the most prominent beam component, and of
these, the dominant helicity state can be chosen by selecting the
sign of the decaying mesons.
The bottom-right panel of Figure
\ref{figure:fluxes} shows a log plot of the energy spectrum of
the NuMI beam in its medium energy (ME) configuration \cite{NuMI}. In
this case, $\nu_\mu$ is the dominant beam component with
$\bar{\nu}_\mu$ contributing at the $3\%$ level and $\nu_e$ making
up less than $1\%$.  We take the NuMI beam in its medium energy,
$\nu_{\mu}$ dominated, configuration (shown in the bottom right panel
of figure \ref{figure:fluxes}) as a representative example
throughout this study.  Optimistically consistent with the above
projections, we assume a $3\%$ overall uncertainty on the total flux
normalization. Similar to the neutrino factory case, the main
sources of background are those events consisting of single
electromagnetic showers (with electron like topologies) which can
generally be removed to a negligible level by cutting on their broad
$p_t$ distribution. Assuming $3.7 \times 10^{20}$ protons on target
(POT) and the given detector configuration, we expect nearly $10^7$
elastic scattering events, inducing a statistical uncertainty of
$0.03\%$.

\setcounter{footnote}{0}
\setcounter{equation}{0}
\section{Results}
\label{sec:Results}

We perform $\chi^2$ analyses to extract the sensitivity of
neutrino--electron elastic scattering experiments to
$\sin^2\theta_W$, $\mu_{\nu}$, $\rho$ and potential leptonic
$Z^{\prime}$-induced couplings $\epsilon$. Table \ref{table:results}
lists the results of our analysis, along with the key assumptions
regarding the different experimental setups. Each result assumes a
one year run with a $100$ ton fiducial mass detector located $100$~m
from the neutrino source.  We take
 $\sin^2\theta_W=0.23120 \pm 0.00015$ at the $Z$-pole for all our analyses, except when
 we explore the ability of the different setups to measure the weak mixing angle itself. We also fix $\rho=1$
 (at the tree-level) for all our analyses, except when we explore the ability of different setups to measure $\rho$
 itself.

\subsection{$\sin^2\theta_W$}

The weak mixing angle $\theta_W$ parameterizes the change of basis
from the $SU(2)_L$ and $U(1)_Y$ gauge fields to the mass
eigenfields, the $W^\pm$-boson, the $Z$-boson, and the photon, after
electroweak symmetry breaking.  Within the SM, electroweak
processes, including those of Eq.~(\ref{eqn:dsig_dT_SM}), can be
expressed in terms of $\sin^2\theta_W$, $G_\mu$ and the
fine-structure constant.  It is essential to precisely measure these
quantities and check for consistency between the various classes of
processes.  The current best fit value is $\sin^2\theta_W(M_Z)  =
0.23120 \pm 0.00015$ \cite{pdg},  or
$\delta(\sin^2\theta_W)/\sin^2\theta_W = 0.065\%$.

The NuTeV collaboration -- which studied deep inelastic
neutrino--nucleus scattering -- extracted a value for
$\sin^2\theta_W$, with $\delta(\sin^2\theta_W)/\sin^2\theta_W =
0.70\%$ precision, that was approximately three standard deviations
above the SM prediction \cite{nutev_results}.  Many possible
explanations of this discrepancy have been proposed
\cite{nutev_explain}, but additional precision measurements must be
made to help pinpoint the true culprit, be it new physics or some
subtle systematic effect.  In particular, neutrino--electron
scattering experiments, utilizing the sources described in
Sec.~\ref{sec:fluxes}, should be especially helpful in this
endeavor, as they may be subject to the same new phenomena
responsible for the NuTeV result, without hadronic complications. It
is in this spirit that we summarize the existing (and proposed)
measurements of the weak mixing angle via neutrino electron
scattering and present our results.

The most precise ($\delta \sin^2\theta_W/\sin^2\theta_W = 0.069\%$)
measurements of the weak mixing angle were done at $e^+e^-$
colliders operating near the $Z$-pole and dominated by the LEP and
SLD experiments \cite{LEP_SLD}.  Compared with such precision
measurements, the past contribution from neutrino--electron elastic
scattering is quite feeble\footnote{Although measurements of the
weak mixing angle are much less precise at neutrino--electron
scattering experiments, their contributions are still very
important.  The variability of the various neutrino beam energies
help to demonstrate the running of $\sin^2\theta_W$.  Additionally,
such processes aid in the search for new physics by signaling
inconsistencies with the $Z$-pole results.} at $\delta
\sin^2\theta_W/\sin^2\theta_W = 3.5\%$ \cite{pdg} resulting mainly
from data taken with the CHARM II detector at the CERN SPS
\cite{charm2_s}, and to a lesser degree from the E734 experiment at
the Brookhaven National Laboratory \cite{E734_Brookhaven}.  Both
experiments, performed with conventional beams, analyzed the ratio
$R = \sigma(\nu_\mu e)/\sigma(\bar{\nu}_\mu e)$ in order to exploit
the cancelation of common systematic uncertainties.  They therefore
took advantage of the ability to run their respective beams in
neutrino/anti-neutrino mode at will, a method that we do not explore
in our analyses.

Our results on the weak mixing angle are as follows:  At a future
reactor experiment, one should be able to measure $\sin^2\theta_W$ with a
$0.82\%$ uncertainty, a result consistent with an estimate made, under similar
assumptions, in  \cite{Conrad_reactor}.  A
neutrino factory experiment, running in either $\mu^+$ or $\mu^-$
mode, can do much better in the absence of systematic uncertainties.
Assuming $0\%$($5\%$) systematic
uncertainty $\delta(\sin^2\theta_W)/\sin^2\theta_W$ could reach $0.14\%$($6.64\%$)
and $0.04\%$($8.62\%$) at a $\mu^+$ and $\mu^-$ 50~GeV
neutrino factory, respectively.  These results are consistent with
the estimates of  \cite{front_end_nufact} (which assume a smaller
detector situated slightly closer to the source). Measurements of
$\sin^2\theta_W$ at low energy ($\gamma < 20$) $\beta$-beam sources have
recently been discussed in  \cite{bb_electroweak} as a function of the expectations for the
systematic uncertainties and the number of different combined boost-factors $\gamma$ used in the
analysis.  They conclude that a $10\%$ measurement of $\sin^2\theta_W$
at low $q^2$ is within reach of a future $\beta$-beam facility,
provided systematic uncertainties are held below $10\%$. Our
analysis assumes a much higher energy ($\gamma = 500$) beam, which
implies larger statistics.  Assuming $0\%$($5\%$) systematic
uncertainties $\delta(\sin^2\theta_W)$ should reach $0.34\%$($7.60\%$)
and $0.22\%$($5.72\%$) at a $\nu_e$ and $\bar{\nu}_e$ $\beta$-beam
respectively.  Finally, at existing or planned conventional neutrino beams,
the weak mixing angle could also be measured.
Using the NuMI beam, we find that $\sin^2\theta_W$ can be
measured with  $0.48\%$($9.92\%$) precision, assuming $0\%$($5\%$)
systematic uncertainty.

\subsection{Neutrino Magnetic Moments}

Neutrino masses imply that neutrinos necessarily have non-zero electromagnetic dipole moments.
The nature of $\mu_{\nu}$ will depend on whether the neutrinos are Majorana or Dirac fermions and, without loss of generality, these are described by (after electroweak simmetry breaking)
\begin{equation}
{\cal L}=\mu_{\nu}^{ij}(\nu_i\sigma_{\mu\nu}\nu_jF^{\mu\nu})+h.c.~~({\rm Majorana}),~~~{\rm or}~~~
{\cal L}=\mu_{\nu}^{ij}(\bar{\nu}_i\sigma_{\mu\nu}\nu_j F^{\mu\nu}) + h.c.~~({\rm Dirac}),
\label{eqn:eff_emLagrangian}
\end{equation}
where $F^{\mu\nu}$ is the electromagnetic field strength.
$\mu_{\nu}^{ij}$ is, in general, complex, and hence carries
information concerning the neutrino electric and magnetic dipole
moments. It will become clear, however, that simply by studying
neutrino--electron scattering it is impossible to decide whether a
non-trivial effect due to Eq.~(\ref{eqn:eff_emLagrangian}) is to be
translated into an electric or magnetic neutrino dipole moment.

\begin{figure}[t]
\begin{center}
\includegraphics[scale=.70]{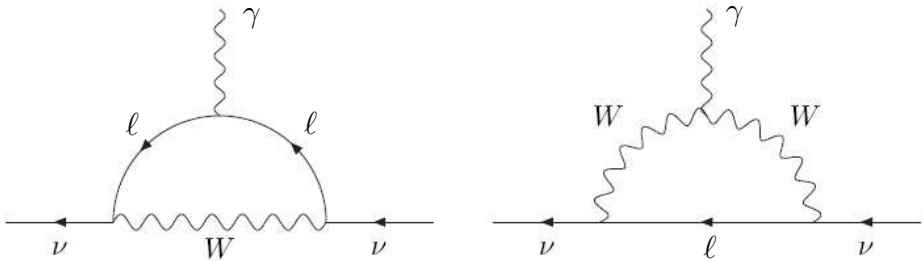}
\caption{Standard Model one-loop contributions to the neutrino magnetic moment.}\label{diag:mmoment}
\end{center}
\end{figure}

In the SM, a  non-zero neutrino magnetic moment is generated at the one-loop level
through the electroweak diagrams depicted in Fig.~\ref{diag:mmoment} and is given, in terms of the
Bohr magneton $\mu_B = e/2m_e$, by \cite{SM_MagMom}
\begin{equation}
\mu_{\nu}^{ij}\leq
\frac{3eG_F}{8\sqrt{2}\pi^2}m_{\nu}=3\times 10^{-20}\mu_B\left(\frac{m_{\nu}}{10^{-1}~\rm eV}\right).
\end{equation}
This is over eight orders of magnitude below the sensitivity of foreseeable future probes of neutrino magnetic moments. For completeness, we mention that, also in the SM, neutrinos are expected
to have a non-zero electric dipole  moment, which is many, many orders of magnitude smaller than
the SM expectation for the neutrino magnetic dipole moment.
Many manifestations of physics beyond the SM, however, predict much larger values for the neutrino magnetic moment \cite{nu_MM_review}.  The observation of a neutrino
magnetic moment any time in the foreseeable  future implies the existence of physics beyond the standard electroweak interactions.

The $i=j$ elements of $\mu_{\nu}^{ij}$, or \emph{diagonal moments},
couple neutrinos of the same mass, while the $ij$ elements, or
\emph{transition moments}, couple different mass
eigenstates.\footnote{In this discussion we are assuming implicitly
that the $\mu_{\nu}^{ij}$ are expressed in the mass eigenbasis and
define transition and diagonal moments using this convention. In the
flavor basis, the magnetic moment matrix is transformed by the
unitary neutrino mixing matrix $U$. In the Majorana case, for
example, these would be related via
$\mu^{\alpha\beta}=U\mu^{ij}U^{T}$.} In the case of Majorana
neutrinos,  $\mu_{\nu}^{ij}$ is constrained to be anti-symmetric by
CPT invariance; that is the requirement that the neutrino and
anti-neutrino must have magnetic moments of equal magnitude.  Thus,
Majorana neutrinos possess only transition moments, whereas Dirac
neutrinos can possess both diagonal and transition moments (this
statement is  weak-basis independent).

The presence of Eq.~(\ref{eqn:eff_emLagrangian})
modifies the neutrino--electron elastic scattering cross-sections
given by Eq.~(\ref{eqn:dsig_dT_SM}) in a dramatic way.  The
important point to realize is that the final state in this process
contains a right-handed neutrino state; fundamentally distinguishable from
the SM final state contribution. It must therefore be
added incoherently to the SM event rate.  Furthermore, a
calculation of the total effect requires a sum over all possible
final state neutrinos since they are unobserved. The net result is
the additional term in the cross-section:
\begin{equation}
\frac{d\sigma}{dT}(\nu_j e \rightarrow \nu_i e)_{em} =
\mu^2\frac{\pi \alpha^2}{E_\nu m_e^2}\left(\frac{E_\nu}{T} - 1
\right). \label{eqn:mmoment_cross}
\end{equation}
The signature of the magnetic moment effect is therefore an excess
of events above the SM prediction displaying a
characteristic $T^{-1}$ dependance.  Here, $\mu$ is now an
effective dipole moment, generally given by $\mu_j^2 = \sum_i|\mu_{\nu}^{ij}|^2$,
where the sum is performed over all possible final states. Of course,
in the case of a short-baseline scattering experiment, the incoming neutrino
is best represented by a flavor eigenstate, in which case it is most practical
to constrain $\mu_{\alpha}=\sum_{\beta}\mu^{\alpha\beta}$, $\alpha,\beta=e,\mu,\tau$. Notice that this
makes it impossible to determine the nature of the neutrino in such
scattering experiments, as one cannot distinguish transition from
diagonal moments.  Matters are complicated further when dealing with
sources composed of multiple neutrino flavors.  In this case one
experimentally extracts (or places an upper limit on) an effective
moment, which is a weighted average of the moments of each beam
flavor component: $\mu^2_{{\rm eff}} = \sum_\alpha \mu^2_\alpha
f_\alpha$, where
\begin{equation}
\frac{d f_\alpha(T)}{d T} = \frac{ \int_T^\infty dE_\nu \frac{d
\Phi_\alpha(E_\nu)}{d E_\nu} \left(\frac{E_\nu}{T} - 1\right)}{\sum
_\beta \int_T^\infty dE_\nu \frac{d \Phi_\beta(E_\nu)}{d E_\nu}
\left(\frac{E_\nu}{T} - 1\right)}.
\end{equation}
$\mu_{{\rm eff}}$ traces out an ellipsoid in magnetic moment space.
If an upper limit $u$ is extracted on the effective moment, the
strongest limit one can place on $\mu_\alpha$ is $\mu_\alpha <
u/\sqrt{f_\alpha}$.  At a 50~GeV neutrino factory, $\mu_{{\rm eff}}$
is related to $\mu_\alpha$  by  $\mu^2_{{\rm eff}} \approx
\frac{6}{11}\mu^2_e + \frac{5}{11} \mu^2_\mu$.  This relation was
obtained by performing the necessary integrals and neglecting all
terms except those proportional to $\ln T^{min}/E_\nu^{max}$.
Notice that both moments contribute about equally, but a slightly
tighter limit can be placed on $\mu_e$ as expected from the low
energy behavior of the neutrino factory energy spectrum (see
Fig.~\ref{figure:fluxes}).  With all of this in mind, it is clear
that the experimentally measured magnetic moment is a very
convoluted quantity, and therefore, one must take great care in
interpreting/comparing such experimental results.

Currently, the tightest bounds on the neutrino magnetic moment comes
from astrophysics \cite{astro_Mmom}.  These limits arise from
considerations of stellar/supernova cooling and are somewhat model
dependent.  Generally, such bounds are of order
$(10^{-10}-10^{-12})\mu_B$, a far cry from the $10^{-19}\mu_B$
predicted from the minimally extended SM.  Direct
measurements via neutrino scattering are less model dependent,
easier to interpret, and are quickly approaching a precision
competitive with astrophysics.

Nuclear reactors offer an ideal setting for studying $\mu_e$ with
$\bar{\nu}_e$ electron scattering due to the low energy peak of the
neutrino spectrum (where magnetic moment effects are most prominent),
and the ability to compare the on/off reactor states.
Most recently, the MUNU \cite{munu} experiment at the Bugey reactor in France
and TEXONO \cite{texono} at the Kuo-Sheng reactor in Taiwan
have analyzed the recoil electron energy spectrum  ${\rm d}N/{\rm d}T$
for very small recoil kinetic energies, $T\lesssim 1$~MeV.
The large uncertainties associated with the
flux normalization at these energies were overcome by the potentially
huge magnetic moment-induced excess that would either dwarf
the SM background or allow the extraction of a strong
upper bound.  Their respective $90\%$ confidence limits are $9\times
10^{-11}\mu_B$ and $1.3\times 10^{-10}\mu_B$.  Our analysis, on the other hand, yields
a 68\% confidence level upper bound of  of $4.8\times 10^{-10}\mu_B$ by
considering the high energy tail of the spectrum, where the overall
normalization is well-known.  $\mu_e$ is also accessible at beta
beam sources.  A 68\% upper bound of $3.0(6.6)\times 10^{-10}\mu_B$ and
$2.6(6.7)\times 10^{-10}\mu_B$ is achievable at a $\nu_e$,
$\bar{\nu}_e$ $\beta$-beam source, respectively, assuming $0\%$($5\%$)
systematic uncertainty.

As for the other neutrino flavors: the LSND \cite{LSND_MMom} experiment provides the
best upper limit value of $6.8\times 10^{-10}\mu_B$ on $\mu_\mu$.  A
future neutrino factory experiment could, at best, improve on this by a
factor of $10$.  With a $0\%$($5\%$) assumed systematic uncertainty,
a 50~GeV neutrino factory experiment can produce an upper
bound on the effective magnetic moment of $2.5(10.1)\times
10^{-11}\mu_B$ and $3.1(12.4)\times 10^{-11}\mu_B$ at $68\%$
confidence for a $\mu^+$ and $\mu^-$ beam, respectively.  The
corresponding bounds on $\mu_e$ and $\mu_\mu$ can be found by
including the factors $1.35$ and $1.48$, respectively as described
above.  For the NuMI beam, an upper limit on $\mu_\mu$ of
$1.8(6.6)\times 10^{-10}\mu_B$ can be achieved, and could be pushed
 below $10^{-10}\mu_B$ with a new/upgraded proton
driver \cite{proton_driver}.  Thus far we have said nothing about
$\mu_\tau$.  At the energies considered here $\nu_\tau$'s cannot be
produced, save for $\nu_\mu \rightarrow \nu_\tau$ oscillation
effects which are negligible at our assumed short baseline of 100~m.
Although our analysis cannot add to the subject, we point out, for
completeness, that the Fermilab DONUT experiment has set a weak
upper bound of $3.7\times 10^{-7}\mu_B$ on $\mu_\tau$ at $90\%$
confidence \cite{donut}.

\subsection{Neutrino $Z^{\prime}$ Couplings: $\epsilon$}

The existence of hypothetical heavy states that couple to both electrons and neutrinos
would modify the differential cross-section for neutrino--electron scattering. Such
heavy states are ubiquitous in models of the physics that lies beyond the SM.

At center-of-mass energies well below the masses of the hypothetical
heavy states, contributions to neutrino--electron scattering are
very well-captured by the introduction of effective four-fermion
operators of the type $\bar{\nu}\Gamma\nu' \bar{e}\Gamma e$, where
$\Gamma$ stand for the various distinct combinations of Dirac gamma
matrices, while $\nu$  and $\nu'$ stand for potentially distinct
neutrino flavors. Here, we will concentrate on flavor independent
vector--vector interactions, described by
\begin{equation}
\mathcal{L} = \epsilon \left(2\sqrt{2}G_\mu\right) \bar{\nu}_\ell
\gamma^\alpha \nu_\ell \bar{e} \gamma_\alpha e + h.c.,
\label{eqn:eff_ZpLagrangian}
\end{equation}
where $\epsilon$ is the new coupling constant, and refer to, for
example, \cite{nsni,nsni_e} for a more detailed study. While we
appreciate the fact that this phenomenon is potentially much richer,
for the purposes of our study, computing the sensitivity to
$\epsilon$, as defined in Eq.~(\ref{eqn:eff_ZpLagrangian}), will
suffice in order to estimate the ability of future $\nu
e$-scattering experiments to probe non-standard neutrino
interactions. On the other hand, the presence of a new, heavy
neutral gauge boson (which we refer to, generically, as a $Z'$) that
couples universally to all three neutrinos and to right-handed and
left-handed electrons with equal strength would yield such an
effective Lagrangian. Hence, we refer to $\epsilon$ as the
neutrino--$Z'$ coupling.

  With the addition of Eq.~(\ref{eqn:eff_ZpLagrangian}),
  Eq.~(\ref{eqn:dsig_dT_SM}) still describes $\nu e$ scattering, as long as one replaces
\begin{eqnarray}
a &\rightarrow& a + \epsilon,\\
b &\rightarrow& b + \epsilon. \nonumber
\end{eqnarray}

At a reactor facility, we find that a $68\%$ upper limit of $2\times
10^{-3}$ can be set on $\epsilon$.\footnote{A detailed study of the
electron $g_L$ and $g_R$ couplings using reactor data  was performed
in \cite{Rosner_UST}. Bounds on $g_L$ and $g_R$ can be easily
converted into bounds on $\epsilon$ (among other possibilities).}
The limits set at a neutrino factory are potentially one order of
magnitude better, $6.9(13.1)\times 10^{-4}$ and $3.3(8.7)\times
10^{-4}$ for a $\mu^+$ and $\mu^-$ beam respectively, assuming
$0\%$($5\%$) systematic uncertainty.  Our estimate for the neutrino
factory agree with estimates obtained in \cite{nsni}.  At a
beta-beam facility we find 68\% upper bounds for $\epsilon$ of
$9.8(16.3)\times 10^{-4}$ and $7.7(14.2)\times 10^{-4}$ for the
$\nu_e$ and $\bar{\nu}_e$ modes respectively. Finally, a
conventional neutrino beam should set a bound of $2.7(6.4)\times
10^{-3}$. Other neutrino sources (including the sun) also allow one
to probe for the existence of new neutrino--electron interactions,
as recently discussed in \cite{nsni_e}.

\subsection{Nature of the neutrino--$Z$-boson Coupling: $\rho$}

In the SM, the neutrino coupling to $Z$-bosons is purely left-handed. In Sec.~\ref{sec:formalism},
we defined the left-handed neutrino--$Z$-boson coupling as $\rho$, which is,
in the SM, equal to unity  at tree-level.
Its interesting to appreciate, however, that, experimentally, the left-handed nature of the
neutrino coupling to the $Z$-boson is far from an established
fact (for a detail discussion of this issue, see \cite{InvisibleZ}).

The most precise information regarding the neutrino--$Z$-boson
coupling is provided by precision studies of the invisible $Z$-boson
width \cite{LEP_SLD}. These, however, are not sensitive to the
left-handed neutrino coupling to the $Z$-boson, but to a combination
of the right-handed and the left-handed couplings to the $Z$-boson.
More insight can only be obtained by combining $Z$-pole data with
that obtained in neutrino scattering. The most  robust bound on the
left-handed--$Z$-boson coupling is obtained by combining $Z$-pole
and CHARM II data \cite{Charm2_nuZ}. According to \cite{InvisibleZ},
$\rho$ values as small as 0.9 are not ruled out (at around the three
sigma level) as long as the right-handed neutrino--$Z$-boson
coupling is nonzero (it is currently bound to be roughly less than
40\% of the left-handed one \cite{InvisibleZ}).

The main point is that, in neutrino--electron scattering, the
incoming neutrino (antineutrino) beams are purely
left-handed(right-handed). Hence, regardless of whether there are
right-handed neutrino $Z$-boson couplings, the neutral current
contribution to  $\nu_\ell e \rightarrow \nu_\ell e$ is only
dependent on $\rho$. For the strictly neutral current processes
$\nu_\ell e \rightarrow \nu_\ell e$ and $\bar{\nu_\ell }e
\rightarrow \bar{\nu_\ell} e$ ($\ell=\mu,\tau$),  the cross section
is proportional to $\rho^2$.\footnote{For simplicity, we assume that
$\rho$ is flavor independent.} As only the total event rates are
affected, it is expected that the capability of experiments with
$\nu_\mu(\bar{\nu}_\mu)$ beams to constrain $\rho$ should be
limited. This is  the condition of decay-in-flight conventional
neutrino sources, where typically the electron neutrino beam
component only contributes at the sub-percent level.  Currently, the
most precise determination of $\rho$ comes from the CHARM II
\cite{Charm2_nuZ} collaboration at the CERN SPS conventional beam
source.  Their result, consistent with the SM prediction of $\rho =
1$, is precise to $\delta\rho = 3.4\%$.  In our analysis, assuming
the NuMI beam, we conclude that $\rho$ can be measured to only
$3.3\%$($7.3\%$) assuming $0\%$($5\%$) systematic uncertainty;
clearly comparable to the CHARM II result. Most of the uncertainty
is related to the rather poor knowledge of the overall normalization
of the neutrino flux.

Reactions with $\nu_e(\bar{\nu}_e)$ involve both charged and
neutral current terms, and the interference between them induces
non-trivial changes to the recoil energy spectrum ${\rm d}
N_{SM}(T)/{\rm d}T$ described by Eq.~(\ref{eqn:rate_spectrum}).
Allowing for arbitrary $\rho$ values, the $a,b$ parameters of Eq.~(\ref{eqn:dsig_dT_SM})
read:
\begin{equation}
\left.\begin{aligned}
a &=\rho\left(\frac{1}{2}-\sin^2\theta_W\right) - 1\\
b &=\rho\left(-\sin^2\theta_W\right)
\end{aligned}
\right\} \qquad \nu_e e \rightarrow \nu_e e.
\end{equation}
Again, $a\leftrightarrow b$ for the $\bar{\nu}_e$ process. For this reason, measurements
of $\nu_e-e$ scattering (or $\bar{\nu}_e-e$) are sensitive to $\rho$ (and not $\rho^2$). Furthermore,
not only is the total event rate modified, but so is the energy distribution of the recoil electrons.
Finally, $\nu_e-e$ scattering is also sensitive to the \emph{sign} of $\rho$, {\it i.e.}, it depends on
whether the $W$-boson exchange interferes destructively or constructively with the $Z$-boson exchange
contribution \cite{boris_rosen}. In the SM $\rho$ is positive -- the $Z$-boson and $W$-boson
exchange diagrams interfere destructively.

Several of the experimental set-ups considered here can extract
(sometimes with high confidence) the sign of $\rho$.  Indeed, such a
feat has already been accomplished by early experiments sensitive to
$\nu_e-e$ elastic scattering \cite{sign_rho}. They find agreement
with destructive interference ($\rho>0$) at around the five sigma
level. The reactor neutrino experimental setup considered here
should be able to repeat such a sign-determination using electron
antineutrinos (which has not been accomplished yet), as long as it
can accumulate enough statistics and control the uncertainty on the
normalization of the $\bar{\nu}_e$ flux. We find, for example, that
the future reactor experiments listed in \cite{reactor_future} could
easily determine the sign of $\rho$ within one year of data
collection, provided that systematic uncertainties (including flux
normalization) are held below (25-30)\%. Needless to say,
$\beta$-beams should provide the ultimate tool when it comes to
studying this issue in detail.

When setting bounds on $\rho$, we explicitly assume that it is
positive.  Our estimates are summarized in
Table~\ref{table:results}. We find that a future reactor experiment
should measure $\rho$ to $1.1\%$. At a beta-beam source, this can be
reduced  to $0.39\%$($2.4\%$) and $0.75\%$($3.1\%$) for a $\nu_e$
and $\bar{\nu}_e$ beam respectively, assuming $0\%$($5\%$)
systematic uncertainty. At a neutrino factory, assuming the same
range of systematic uncertainties, we expect a precision of
$0.09\%$($1.2\%$) and $0.06\%$($0.93\%$) for a $\mu^+$ and $\mu^-$
beam respectively. As mentioned earlier, we have assumed that the
value of $\rho$ is flavor universal, so that, in the case of a
neutrino factory, information is obtained from both the $\nu_{\mu}$
and the $\nu_e$ components of the beam.

\setcounter{footnote}{0}
\setcounter{equation}{0}
\section{Concluding Remarks}
\label{sec:end}

Neutrino--electron scattering provides a very clean environment for detailed studies of electroweak
interactions. In principle, one is not only capable of precisely determining the value of SM
parameters, but is also sensitive to physics beyond the SM, including anomalous
neutrino couplings to photons, neutrino and electron couplings to new neutral gauge bosons ($Z$ primes), and
right-handed neutrino neutral currents.

On the negative side, the cross-section for neutrino--electron scattering is tiny. This means that one
needs very large neutrino sources and/or neutrino targets. Moreover, backgrounds related to
neutrino--nucleon scattering, whose cross-section is around three orders of magnitude larger, need to
be seriously suppressed. Finally, competitive precision measurements can only be performed if the
neutrino beams are very well understood (shape and normalization).

It is now clear that, in the foreseeable future, new neutrino
facilities, where the obstacles summarized above can be eliminated,
will become available. The new physics revealed by neutrino
oscillation experiments calls for very intense, very well understood
neutrino sources, and these are currently under serious
consideration. Furthermore, many of these planned facilities will
house ``near detectors,'' for several reasons. The types of set-ups
we are considering qualify as near detectors (not unlike the
Miner$\nu$a experiment, currently being planned as a new detector to
be added to the MINOS near detector).

Here, we have estimated how precisely various observables could be
measured via neutrino--electron elastic scattering at existing (say, the NuMI beam) and
future facilities, including neutrino factories, $\beta$-beams, and next-generation, large detectors located close to
powerful nuclear reactors.  Table~\ref{table:results} summarizes our results,
as well as the assumptions that went into extracting them. For most set-ups, we have quoted expectations in the
case that systematic uncertainties are reduced to negligible levels -- results with relatively large systematic
uncertainties are quoted in Sec.~\ref{sec:Results}. Ultimately, the ``correct'' estimate for systematic effects
will be obtained by the experimental collaborations. We believe, however, that our estimates can be considered
representative of either a typical or  a worst-case scenario.

In summary, it is fair to say that, in the foreseeable future, we
can expect neutrino--electron scattering experiments to contribute,
in a significant way, to our understanding of electroweak
interactions -- and beyond. We urge experimentalists to keep the
possibility of performing precision neutrino--electron scattering
studies when developing next-generation ``near detectors'' for
future neutrino facilities.

\begin{table}
\centering
\caption{Results on the precision of parameter extraction, assuming
  a $100$ ton detector located $100$~m from the neutrino source.  All limits are taken at $68\%$ confidence.
  See text for details \label{table:results}.
}.
\begin{tabular}{|c|c|c||c|c|c|c|} \hline
  & \textbf{Assumptions} & \textbf{Uncertainties} & \textbf{$\sin^2\theta_W$} & \textbf{magnetic moment} & \textbf{$Z^{\prime}$ coupling $\epsilon$}&\textbf{ $\rho$ }\\
  & &\% bkg, \% flux &  \% & 68\% & 68\% & \% \\
  \hline
  \textbf{Reactor} & $3 {\rm GW}$, $3<T<5 {\rm MeV}$  \cite{Conrad_reactor}& $1$, $0.1$ & $0.82$ & $4.8\times10^{-10} \mu_B$& $2.0\times 10^{-3}$ & $1.1$ \\
  \textbf{$\mu^+$ $\nu$-factory} & $50{\rm GeV}$, $10^{20}\frac{{\rm decays}}{{\rm year}}$  \cite{front_end_nufact} & $0$, $0.1$ & $0.14$ & $2.5\times 10^{-11} \mu_B$ & $2.1\times 10^{-3}$ & $0.09$ \\
  \textbf{$\mu^-$ $\nu$-factory} & $50{\rm GeV}$, $10^{20}\frac{{\rm decays}}{{\rm year}}$  \cite{front_end_nufact} & $0$, $0.05$ & $0.04$ & $3.1\times 10^{-11} \mu_B$ & $2.0\times 10^{-3}$ &$0.06$ \\
  \textbf{$\beta$-beam} $\nu_e$ ($^{18}{\rm Ne}$,)  & $\gamma=500$, $1.1\times10^{18} \frac{{\rm decays}}{{\rm yesr}}$ \cite{bbeam_nufact_RD} & $0$, $0.1$ & $0.34$ & $3.0\times 10^{-10}\mu_B$& $9.8\times 10^{-4}$ & $0.39$ \\
  \textbf{$\beta$-beam} $\bar{\nu}_e$ ($^6{\rm He}$) & $\gamma=500$, $2.9\times10^{18} \frac{{\rm decays}}{{\rm yesr}}$ \cite{bbeam_nufact_RD} & $0$, $0.1$ & $0.22$ & $2.6\times 10^{-10}\mu_B$ & $7.7\times 10^{-4}$ & $0.75$ \\
  \textbf{Conventional} & NuMI on-axis $3.7\times 10^{20}$ POT& $0$, $3$ & $0.48$ & $1.8\times 10^{-10}\mu_B$ & $2.7\times 10^{-3}$ & $3.3$ \\ \hline
  \end{tabular}
  \end{table}

\section*{Acknowledgments}

We thank Olga Mena for very useful conversations, for information
regarding the neutrino fluxes from different future neutrino
facilities, and for comments on the manuscript. We are also indebted
to Boris Kayser for fruitful conversations on the neutrino coupling
to the $Z$-boson. This work is sponsored in part by the US
Department of Energy Contract DE-FG02-91ER40684.


\begin{thebibliography}{99}



\bibitem{bbeam_nufact_RD}
  C.~Albright {\it et al.}  [Neutrino Factory/Muon Collider Collaboration],
  physics/0411123.

\bibitem{minerva}
  D.~Drakoulakos {\it et al.}  [Minerva Collaboration],
  hep-ex/0405002.



   \bibitem{pdg} S.~Eidelman {\it et al.}  [Particle Data Group Collaboration],
Phys.\ Lett.\ B {\bf 592}, 1 (2004).


\bibitem{Bahcall_radCorr}
  J.~N.~Bahcall, M.~Kamionkowski and A.~Sirlin,
  Phys.\ Rev.\ D {\bf 51}, 6146 (1995).

\bibitem{Marciano_radCorr}
  W.~J.~Marciano and A.~Sirlin,
  Phys.\ Rev.\ D {\bf 22}, 2695 (1980)
  [Erratum-ibid.\ D {\bf 31}, 213 (1985)].

\bibitem{Sarantakos_radCorr}
  S.~Sarantakos, A.~Sirlin and W.~J.~Marciano,
  Nucl.\ Phys.\ B {\bf 217}, 84 (1983).

\bibitem{practical_radCorr}
  M.~Drees,
DESY-91-045
{\it Based on a talk given at 2nd Workshop on High Energy Physics
Phenomenology, Calcutta, India, Jan 2-15, 1991}

\bibitem{nu_eReview}
  W.~J.~Marciano and Z.~Parsa,
  J.\ Phys.\ G {\bf 29}, 2629 (2003).
See also M.~Passera,
  Phys.\ Rev.\ D {\bf 64}, 113002 (2001).

\bibitem{meas_nue_scat}
 See, for example,  R.~Imlay and G.~J.~VanDalen,
  J.\ Phys.\ G {\bf 29}, 2647 (2003).




\bibitem{reactor_review}
  C.~Bemporad, G.~Gratta and P.~Vogel,
  Rev.\ Mod.\ Phys.\  {\bf 74}, 297 (2002).

\bibitem{low_Ereactor}
  H.~B.~Li and H.~T.~Wong,
  J.\ Phys.\ G {\bf 28}, 1453 (2002).

\bibitem{reactor_param1}
  P.~Vogel and J.~Engel,
  Phys.\ Rev.\ D {\bf 39}, 3378 (1989).

\bibitem{reactor_param2}
  H.~Murayama and A.~Pierce,
  Phys.\ Rev.\ D {\bf 65}, 013012 (2002).

\bibitem{prec_reactor_nueb}
  P.~Huber and T.~Schwetz,
  Phys.\ Rev.\ D {\bf 70}, 053011 (2004).

\bibitem{coherent}
for recent studies related to observing this process, see, for example,
H.T.~Wong, H.B.~Li, J.~Li, Q.~Yue and Z.Y.~Zhou,
  hep-ex/0511001.

\bibitem{Conrad_reactor}
  J.~M.~Conrad, J.~M.~Link and M.~H.~Shaevitz,
  Phys.\ Rev.\ D {\bf 71}, 073013 (2005).


\bibitem{CHOOZ}
  M.~Apollonio {\it et al.}  [CHOOZ Collaboration],
  Phys.\ Lett.\ B {\bf 420}, 397 (1998).

\bibitem{reactor_whitepaper}
 K.~Anderson {\it et al.},
 hep-ex/0402041.

\bibitem{reactor_future}
 M.~Goodman,
 Nucl.\ Phys.\ Proc.\ Suppl.\  {\bf 145}, 186 (2005).

\bibitem{DCHOOZ}
 F.~Ardellier {\it et al.},
 hep-ex/0405032.



\bibitem{nufact_norm}
  S.~Geer,
  Phys.\ Rev.\ D {\bf 57}, 6989 (1998)
  [Erratum-ibid.\ D {\bf 59}, 039903 (1999)].



\bibitem{front_end_nufact}
  M.~L.~Mangano {\it et al.},
  hep-ph/0105155.



\bibitem{high_gamma_bbeams}
  J.~Burguet-Castell, D.~Casper, J.~J.~Gomez-Cadenas, P.~Hernandez and F.~Sanchez,
  Nucl.\ Phys.\ B {\bf 695}, 217 (2004).

\bibitem{k2k}
S.~H.~Ahn {\it et al.}  [K2K Collaboration],
  Phys.\ Lett.\ B {\bf 511}, 178 (2001).



\bibitem{t2k_k2k}
For a recent discussion of the current status of T2K, see  Y.~Oyama,
  hep-ex/0512041.


\bibitem{Finesse}
  L.~Bugel {\it et al.}  [FINeSSE Collaboration],
  hep-ex/0402007.


\bibitem{proton_driver}
  M.~G.~Albrow {\it et al.},
  hep-ex/0509019.



\bibitem{NuMI}
    Olga Mena, private communication.

\bibitem{nutev_results}
  G.~P.~Zeller {\it et al.}  [NuTeV Collaboration],
  Phys.\ Rev.\ Lett.\  {\bf 88}, 091802 (2002)
  [Erratum-ibid.\  {\bf 90}, 239902 (2003)].

\bibitem{nutev_explain}
  S.~Davidson, S.~Forte, P.~Gambino, N.~Rius and A.~Strumia,
  JHEP {\bf 0202}, 037 (2002).

\bibitem{LEP_SLD} S. Schael {\it et al.},
  hep-ex/0509008.



\bibitem{charm2_s}
  D.~Geiregat {\it et al.}  [CHARM-II Collaboration],
  Phys.\ Lett.\ B {\bf 259}, 499 (1991);
P.~Vilain {\it et al.}  [CHARM-II Collaboration],
  Phys.\ Lett.\ B {\bf 281}, 159 (1992).



\bibitem{E734_Brookhaven}
  L.~A.~Ahrens {\it et al.},
  Phys.\ Rev.\ D {\bf 41}, 3297 (1990).

\bibitem{bb_electroweak}
  A.~B.~Balantekin, J.~H.~de Jesus and C.~Volpe,
  hep-ph/0512310.

\bibitem{nu_MM_review} For recent reviews see, for example,
R.N.~Mohapatra {\it et al.},
  hep-ph/0510213;
A.~de Gouv\^ea,
  Nucl.\ Phys.\ Proc.\ Suppl.\  {\bf 143}, 167 (2005).


\bibitem{SM_MagMom}
  K.~Fujikawa and R.~Shrock,
  Phys.\ Rev.\ Lett.\  {\bf 45}, 963 (1980).

\bibitem{astro_Mmom}
  H.~T.~Wong and H.~B.~Li,
  Mod.\ Phys.\ Lett.\ A {\bf 20}, 1103 (2005).



\bibitem{munu}
  Z.~Daraktchieva {\it et al.}  [MUNU Collaboration],
  Phys.\ Lett.\ B {\bf 615}, 153 (2005).




\bibitem{texono}
  H.~B.~Li {\it et al.}  [TEXONO Collaboration],
  Phys.\ Rev.\ Lett.\  {\bf 90}, 131802 (2003).



\bibitem{LSND_MMom}
  L.~B.~Auerbach {\it et al.}  [LSND Collaboration],
  Phys.\ Rev.\ D {\bf 63}, 112001 (2001).

\bibitem{donut}
  R.~Schwienhorst {\it et al.}  [DONUT Collaboration],
  Phys.\ Lett.\ B {\bf 513}, 23 (2001).

\bibitem{nsni}
S.~Davidson, C.~Pe\~na-Garay, N.~Rius and A.~Santamaria,
  JHEP {\bf 0303}, 011 (2003).

  \bibitem{nsni_e}
  J.~Barranco, O.G.~Miranda, C.A.~Moura and J.W.F.~Valle,
  hep-ph/0512195.

\bibitem{Rosner_UST}
  J.~L.~Rosner,
  Phys.\ Rev.\ D {\bf 70}, 037301 (2004).



\bibitem{InvisibleZ}
  M.~Carena, A.~de Gouv\^ea, A.~Freitas and M.~Schmitt,
  Phys.\ Rev.\ D {\bf 68}, 113007 (2003).

\bibitem{Charm2_nuZ}
  P.~Vilain {\it et al.}  [CHARM-II Collaboration],
  Phys.\ Lett.\ B {\bf 320}, 203 (1994).

\bibitem{boris_rosen}
B.~Kayser, E.~Fischbach, S.P.~Rosen and H.~Spivack,
  Phys.\ Rev.\ D {\bf 20}, 87 (1979).

\bibitem{sign_rho}
  R.C.~Allen {\it et al.},
  Phys.\ Rev.\ Lett.\  {\bf 64}, 1330 (1990);
L.B.~Auerbach {\it et al.}  [LSND Collaboration],
  Phys.\ Rev.\ D {\bf 63}, 112001 (2001).





 \end{thebibliography}
 \end{document}